\documentclass[aps,twocolumn,reprint,amsmath,amssymb,showpacs,preprintnumbers]{revtex4-1}
\usepackage{graphicx}
\usepackage{dcolumn}
\usepackage{bm}
\usepackage{hyperref}
\usepackage{textcomp} 
\usepackage{color}

\begin{document}

\title{Onsager vortex formation in Bose--Einstein condensates in two-dimensional \protect\\ power-law traps}
\author{Andrew J. Groszek, Tapio P. Simula, David M. Paganin and Kristian Helmerson \\	
\textit{School of Physics and Astronomy, Monash University, Victoria 3800, Australia}}
	
\begin{abstract}

We study computationally dynamics of quantised vortices in two-dimensional superfluid Bose--Einstein condensates confined in highly oblate power-law traps. We have found that the formation of large scale Onsager vortex clusters prevalent in steep-walled traps is suppressed in condensates confined by harmonic potentials. However, the shape of the trapping potential does not appear to adversely affect the evaporative heating efficiency of the vortex gas. Instead, the suppression of Onsager vortex formation in harmonic traps can be understood in terms of the energy of the vortex configurations. Furthermore, we find that the vortex--antivortex pair annihilation that underpins the vortex evaporative heating mechanism requires the interaction of at least three vortices. We conclude that experimental observation of Onsager vortices should be the most apparent in flat or inverted-bottom traps.

\end{abstract}

\preprint{DOI: \href{http://dx.doi.org/10.1103/PhysRevA.93.043614}{10.1103/PhysRevA.93.043614}}

\maketitle


\section{Introduction \label{sec:intro}}

Non-equilibrium physics of quantum gases has attracted significant activity recently \cite{giamarchi_strongly_2012}. Quantum turbulence (QT) is an archetype of non-equilibrium dynamics, which manifests as a chaotic motion of large numbers of quantised vortices and features an intriguing interplay between chaos and order. In three-dimensional QT, vortex filaments form tangles---a phenomenon which has been widely studied but only imaged directly in recent years in superfluid helium \cite{bewley_superfluid_2006, paoletti_velocity_2008, guo_visualization_2014}. Remarkably, despite the fact that the microscopic behaviour of three-dimensional QT is driven by Kelvin waves
 \cite{thomson_xxiv._1880,Bretin2003a,Simula2008a,koens_vibrations_2013,Fonda2014a},
 Crow instabilities \cite{Crow1970a,Berloff2001a,Simula2011a}, vortex reconnections \cite{Feynman1955a,Schwarz1985a,Fonda2014a}, phonon radiation \cite{samuels_vortex_1998, barenghi_decay_2005} and mutual friction between the normal and superfluid components \cite{hall_rotation_1956}, statistically the dynamics is thought to yield the same Kolmogorov scaling of incompressible kinetic energy as in classical fluid turbulence.

Restricting the motion of the quantised vortices in one of the spatial dimensions results in two-dimensional (2D) quantum turbulence, in which the vortex tangle reduces to a chaotic configuration of point-like vortices. Recent studies have focused on observing the decay of QT in Bose--Einstein condensates, both experimentally \cite{neely_characteristics_2013, kwon_relaxation_2014, seman_route_2011} and using computer simulations \cite{parker_emergence_2005,nazarenko_freely_2006, numasato_possibility_2009, numasato_direct_2010,schole_critical_2012,reeves_inverse_2013, simula_emergence_2014, billam_onsager-kraichnan_2014, billam_spectral_2015, stagg_generation_2015}. Two-dimensional systems have attracted particular interest due to a prediction of an inverse energy cascade \cite{kraichnan_inertial_1967,kraichnan_inertial-range_1971} from small to large spatial scales, which originates from the theory of classical fluid turbulence. Using a statistical model of point-vortices, Onsager \cite{onsager_statistical_1949} predicted for such systems the emergence of large-scale vortex structures, such as those seen in geophysical systems \cite{eyink_onsager_2006}. Similarly, in 2D QT, the inverse cascade is anticipated to lead to the clustering of like-sign vortices into large-scale Onsager vortex structures.

Recent experimental advances in producing \cite{seman_route_2011, wilson_experimental_2013, kwon_critical_2015, kang_rotating_2015}, imaging \cite{wilson_situ_2015} and controlling \cite{samson_deterministic_2016} quantised vortices in ultracold atomic dilute gas Bose--Einstein condensates (BECs) have resulted in detailed measurements of vortex dynamics in these superfluid systems \cite{freilich_real-time_2010, kwon_relaxation_2014, navarro_dynamics_2013, neely_observation_2010}.  However, the debate continues regarding whether or not an inverse cascade and associated Onsager vortices should emerge in compressible 2D QT \cite{neely_characteristics_2013, billam_onsager-kraichnan_2014, billam_spectral_2015, reeves_inverse_2013, simula_emergence_2014, reeves_inverse_2013, chesler_vortex_2014, kwon_relaxation_2014, stagg_generation_2015, numasato_direct_2010, white_creation_2012}. Numasato \emph{et al.}~\cite{numasato_direct_2010} simulated quantum turbulence in a uniform 2D superfluid and found evidence of a direct cascade pushing incompressible kinetic energy towards small length scales. In accordance with this finding, a recent experiment \cite{kwon_relaxation_2014} and simulation \cite{stagg_generation_2015} of a turbulent harmonically trapped highly oblate BEC did not find evidence for the formation of Onsager vortices. By contrast, Simula \emph{et al.}~\cite{simula_emergence_2014} observed strong evidence of vortex clustering in their quasi-2D simulations in a flat trap with steep walls.

One key difference between these studies which could explain the disparity between their findings is the trapping potential used for confining the condensate. The aim of this paper is therefore to investigate the role of the trap geometry with regard to the emergence of Onsager vortices. We focus on numerical studies of decaying two-dimensional quantum turbulence in power-law traps, with a particular emphasis on comparing harmonically trapped condensates to those in uniform disk potentials with steep walls. A variety of techniques exist for producing such steep-walled trapping potentials experimentally \cite{gaunt_bose-einstein_2013,Chomaz2015a,Lee2015a}.

We simulate BEC dynamics using a Gross--Pitaevskii model and also study their thermodynamic properties using a Markov Chain Monte Carlo technique, interpreting the vortex dynamics in each trap in terms of vortex evaporative heating \cite{simula_emergence_2014}. In addition, we examine in detail the microscopic process of vortex--antivortex annihilation, an essential aspect of the decaying turbulence in these systems. In Sec.~\ref{sec:model}, we introduce the numerical model and computational techniques. In Sec.~\ref{sec:results}, we present the key findings from our simulations of decaying superfluid turbulence in different trapping potentials and interpret our observations using a statistical mechanics framework. We then examine the vortex dynamics on a microscopic scale, focusing in particular on vortex--antivortex annihilation in 2D QT, showing it to be a many-vortex process. Finally, Sec.~\ref{sec:discussion} is devoted to discussion.

\section{Model \label{sec:model}}

\subsection{System parameters \label{sub:systemparams}}
The vortex dynamics of two-dimensional dilute gas Bose--Einstein condensates are inherent in the time-dependent condensate wavefunction $\psi(\textbf{r},t)$, whose evolution is modelled here using the Gross--Pitaevskii equation (GPE):
\begin{equation} \label{eq:GPE}
i \hbar \frac{\partial}{\partial t} \psi(\textbf{r},t) = \left[\frac{-\hbar^2}{2m}\nabla^2 + V_{\textrm{trap}}(\textbf{r}) + g_{\rm 2D}|\psi(\textbf{r},t)|^2\right]\psi(\textbf{r},t),
\end{equation}
where $m$ is the mass of an atom, and $V_{\textrm{trap}}(\textbf{r})$ is the trapping potential which radially confines the condensate. The effective interaction parameter $g_{\rm 2D} = g N \int |\psi_z(z)|^4 dz$ accounts for reduction of dimensionality of the Gross--Pitaevskii equation from three to two, where $\psi_z$ is the normalised axial wavefunction, $N$ is the total number of atoms in the condensate, and $g = 4 \pi \hbar^2 a_s / m$ is the interaction parameter for the three-dimensional system being modeled, defined in terms of the $s$-wave scattering length $a_s$. For a uniform cylindrical trap, $g_{\rm 2D} = g N / l_z$, where $l_z$ is the axial length of the three-dimensional condensate. In a sufficiently tight axial harmonic trap, the axial wave function is well approximated by a Gaussian, and hence $g_{\rm 2D} = g N / \sqrt{2 \pi} a_z$, where $a_z = \sqrt{\hbar/m\omega_z}$ is the axial harmonic oscillator length with an effective harmonic trapping frequency $\omega_z$. The wavefunction $\psi_{\rm 3D}({\bf r},z)=\psi({\bf r})\psi(z)$ is normalized such that $\int |\psi({\bf r})|^2 dxdy =  \int |\psi(z)|^2 dz = 1$.

We consider general power-law trapping potentials
\begin{equation} \label{eq:Vtrap}
V_{\textrm{trap}}(\textbf{r}) = \frac{1}{2} m \omega_r^2 R_\textrm{o}^2 \left( \frac{|\textbf{r}|}{R_\textrm{o}} \right) ^{\alpha},
\end{equation}
where $\omega_r$ is the radial harmonic trapping frequency, $\alpha$ is a parameter which defines the steepness of the trap walls, and $R_\textrm{o}$ is the effective system radius. For $\alpha=2$ this potential is a standard harmonic trap $V(\textbf{r}) = \frac{1}{2} m \omega_r^2 |\textbf{r}|^2$ with Thomas--Fermi radius $R_{\textrm{TF}} =R_\textrm{o}$. In the limit of infinite steepness ($\alpha \rightarrow \infty$) it approaches a cylindrically symmetric well of radius $R_\textrm{o}$.

Our system parameters correspond to a two-dimensional $^{23}\textrm{Na}$ BEC with a radial trapping frequency of $\omega_{r}=2\pi \times 15\,\textrm{Hz}$, and a Thomas--Fermi radius of $R_{\textrm{TF}} \approx 70 \, \mu \textrm{m} \approx 12.79 \, a_{\textrm{osc}}$, where the radial harmonic oscillator length scale is defined as $a_{\textrm{osc}} = \sqrt{\hbar / m \omega_r}$. To this end we choose $g_{\rm 2D}=21000\,\hbar^2 / m$. Hence, the radial extent of our system is similar to those used in the recent experiment by Kwon \emph{et al.}~\cite{kwon_relaxation_2014} and simulations by Stagg \emph{et al.}~\cite{stagg_generation_2015}. 

\subsection{Numerical techniques \label{sub:techniques}}

In the beginning of our simulations we solve for the approximate ground state of the system using imaginary time evolution of the GPE. We then imprint $N_v(t=0)=120$ vortices in the condensate by multiplying the ground state wavefunction by a phase factor $\prod_k^{N_v} \textrm{exp}(i \theta_k)$, where $\theta_k(x,y) = s_k \textrm{arctan}[(y - y_k) / (x - x_k)]$. Here, the co-ordinate $(x_k,y_k)$ defines the position of the $k$th vortex, whose circulation sign is $s_k$. We imprint equal numbers of vortices ($s_k=1$) and antivortices ($s_k=-1$).

We choose initial conditions which approximate high entropy, highly randomised states which could be produced by stirring the condensate. To this end, we first construct a density of states distribution $D(E)$ for our chosen vortex number by iteratively generating random vortex configurations and calculating their energy $E$ using a point-vortex Hamiltonian \cite{simula_emergence_2014}. The maximum entropy state corresponds to the peak of this distribution; hence, we ensure that all initial conditions generated have an energy lying within 10\% of this maximum entropy value \footnote{The evaporative heating mechanism does not rely on starting with a specific vortex configuration---the initial condition simply determines how much heating is required to reach the clustered Onsager vortex states.}. 

After the vortex imprinting step, the wavefunction is evolved further in imaginary time for $0.05\,\omega_r^{-1}$ to establish the structure of the vortex cores. This can lead to the annihilation of vortices near the boundary, as well as vortex--antivortex pairs if they were imprinted very close together. The number of vortices at the start of the real-time evolution is therefore on average seven fewer than the $120$ that were originally imprinted.

We solve the GPE using a fourth-order split-step Fourier method on a $1024 \times 1024$ spatial grid with spacing $ \Delta x \approx 0.05\,a_{\textrm{osc}}$ (approximately 0.65 condensate healing lengths) unless otherwise stated. The locations of the vortices in the system are detected by measuring the positions of the phase singularities in the wavefunction at predetermined time intervals. The direction of the phase winding about each singularity determines the circulation sign of the vortex. The vortex locations are only measured in a region $|\textbf{r}|<0.9 \, R_\textrm{o}$ in order to avoid detection of ghost vortices \cite{tsubota_vortex_2002} in the low density region of the traps with lower $\alpha$ values.


\section{Results \label{sec:results}}

\subsection{Macroscopic dynamical behaviour \label{sub:macro}}

We first compare the results of decaying turbulence in the two traps discussed in the literature: a harmonic trap ($\alpha=2$) and a uniform trap with steep walls ($\alpha=100$), which has constant density to within $\sim 5$ healing lengths of the boundary. For each simulation, we monitor the number of vortices $N_v(t)$, which decreases over time due to vortex annihilation events. We also measure the dipole moment $d(t)$ of the vortex distribution, defined as {$d = \left| \textbf{d} \right| = \left| \sum _i q_i \textbf{r}_i \right|$, where $\textbf{r}_i$ is the position of the \textit{i}th vortex, and $q_i = s_i \kappa = s_i h / m$ is its charge. For the confined systems being studied here, it is convenient to scale $d$ with the system size $R_{\rm o}$ and the number of vortices $N_v$. If the vortices are randomly distributed, $d$ will approach zero for large systems. A large $d$, on the other hand, signals the presence of two Onsager vortex clusters in our system.

Figure~\ref{fig:dynamics} shows the characteristic time evolution of the vortex distribution in the two traps, along with the respective dipole moments. In agreement with previous simulations and experiments \cite{stagg_generation_2015, kwon_relaxation_2014}, we observe no significant vortex clustering in the $\alpha=2$ harmonic trap. However, and also in agreement with previous 3D simulations \cite{simula_emergence_2014}, the uniform trap exhibits a strong tendency to form Onsager vortices, as indicated by the increasing dipole moment. Thus, we conclude that the shape of the trapping potential has a strong influence on the vortex clustering behaviour, partially resolving the apparent contradiction in the existing literature.

\begin{figure}[t]
\centering
\includegraphics[width=\columnwidth]{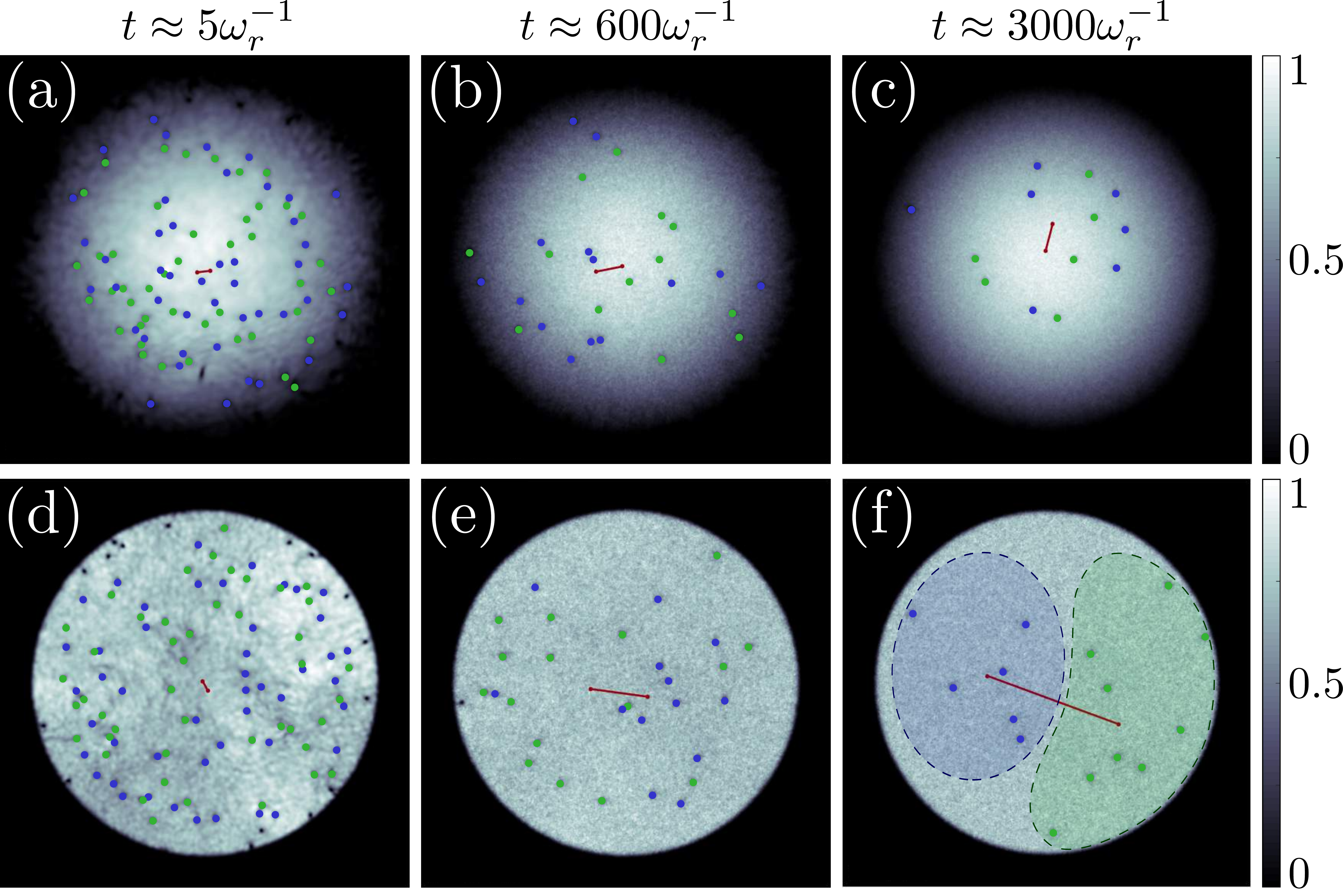}
\caption{Comparison of the time evolution of the vortex configuration between the $\alpha=2$ harmonic trap (a)--(c) and $\alpha=100$ uniform trap (d)--(f). The grayscale value represents the superfluid density, and the colorbars are normalised to the maximum density: $4.3 \times 10^{-3} \, a_{\rm{osc}}^{-2}$ and $2.7 \times 10^{-3} \, a_{\rm{osc}}^{-2}$ for the top and bottom rows, respectively. Vortices and antivortices are denoted by blue (dark) and green (light) circles, respectively. The red line represents the effective dipole moment of the vortex distribution. Movies S1 and S2 in the Supplemental Materials \cite{supplement} show the dynamics of each simulation.}
\label{fig:dynamics}
\end{figure}


\subsection{Statistical mechanics interpretation \label{sub:statinterp}}

The spontaneous formation of Onsager vortices found in Ref.~\cite{simula_emergence_2014} was attributed to the evaporative heating mechanism of vortices.  When a vortex--antivortex pair annihilates, the incompressible kinetic energy $E_{\textrm{inc}}$ of the system is redistributed among the vortices remaining in the system. This process can lead to evaporative heating of the vortex gas, whereby the mean energy per vortex increases each time an annihilation occurs. When the mean energy per vortex crosses a critical value, a transition into the Onsager vortex state is possible \cite{simula_emergence_2014}.


\begin{figure}[t]
\includegraphics[width=\columnwidth]{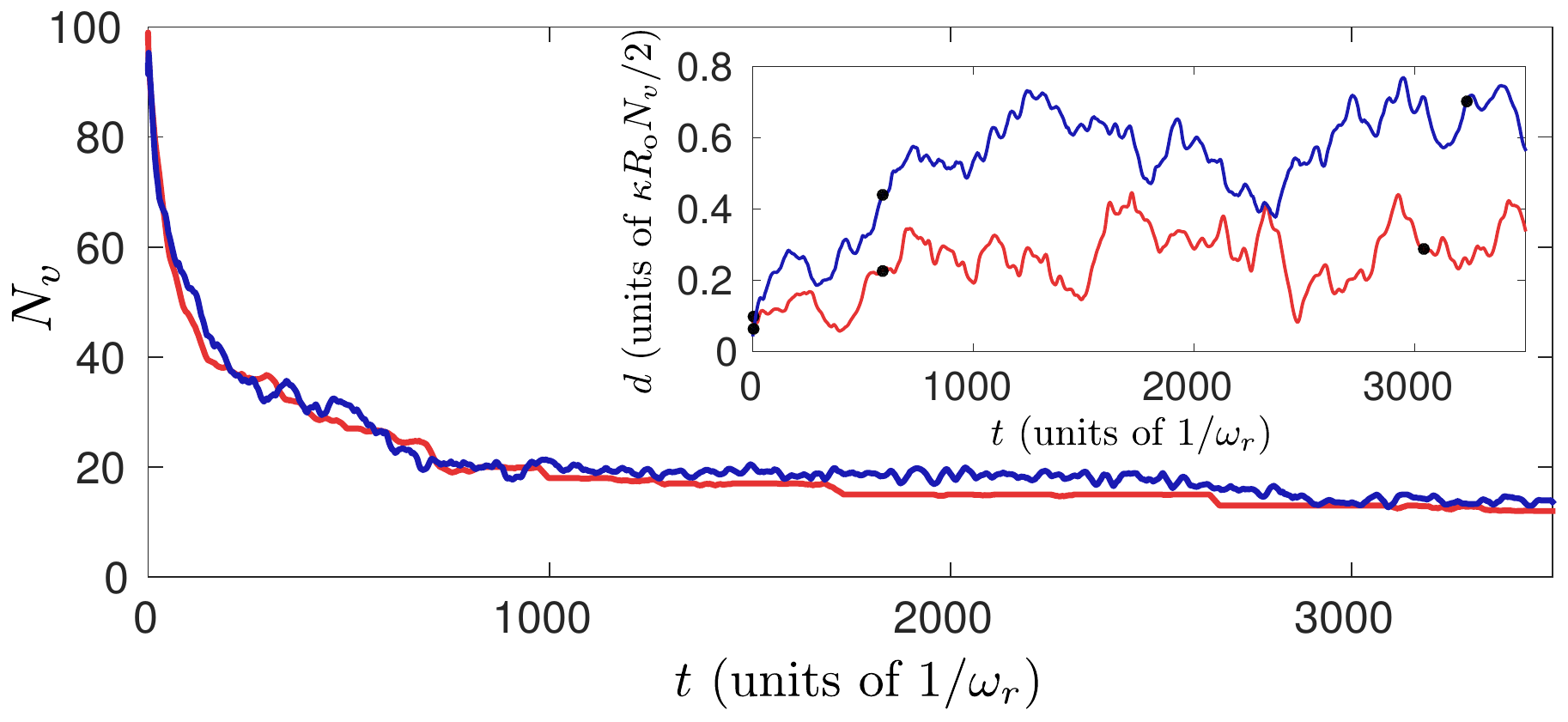}
\caption{Comparison of the vortex number decay and dipole moment evolution (inset) for the harmonic (red/light) and uniform (blue/dark) traps. The black circles in the inset correspond to the timeframes displayed in Fig.~\ref{fig:dynamics}. The fluctuations in the vortex number are due to vortices crossing the counting radius of $0.9\, R_{\textrm{o}}$, in addition to occasional vortex--antivortex pair creation.}
\label{fig:stats}
\end{figure}

The absence of strong clustering in the harmonic trap could be due to (i) the rate of evaporative heating per annihilation event being too low, leading to inefficient evaporative heating of the vortex gas, (ii) the critical energy per vortex for the Onsager vortex transition in a harmonic trap being out of reach despite the vortices being evaporatively heated, or (iii) the critical value of the dipole moment for harmonic traps being too small to allow a clear distinction to be made between the disordered and clustered vortex configurations. In the following we argue that the combined effect of (ii) and (iii) may explain the observed behaviour.

\subsubsection{Dynamical statistical behaviour \label{ssub:stats1}}

\begin{figure}[b]
\centering
\includegraphics[width=\columnwidth]{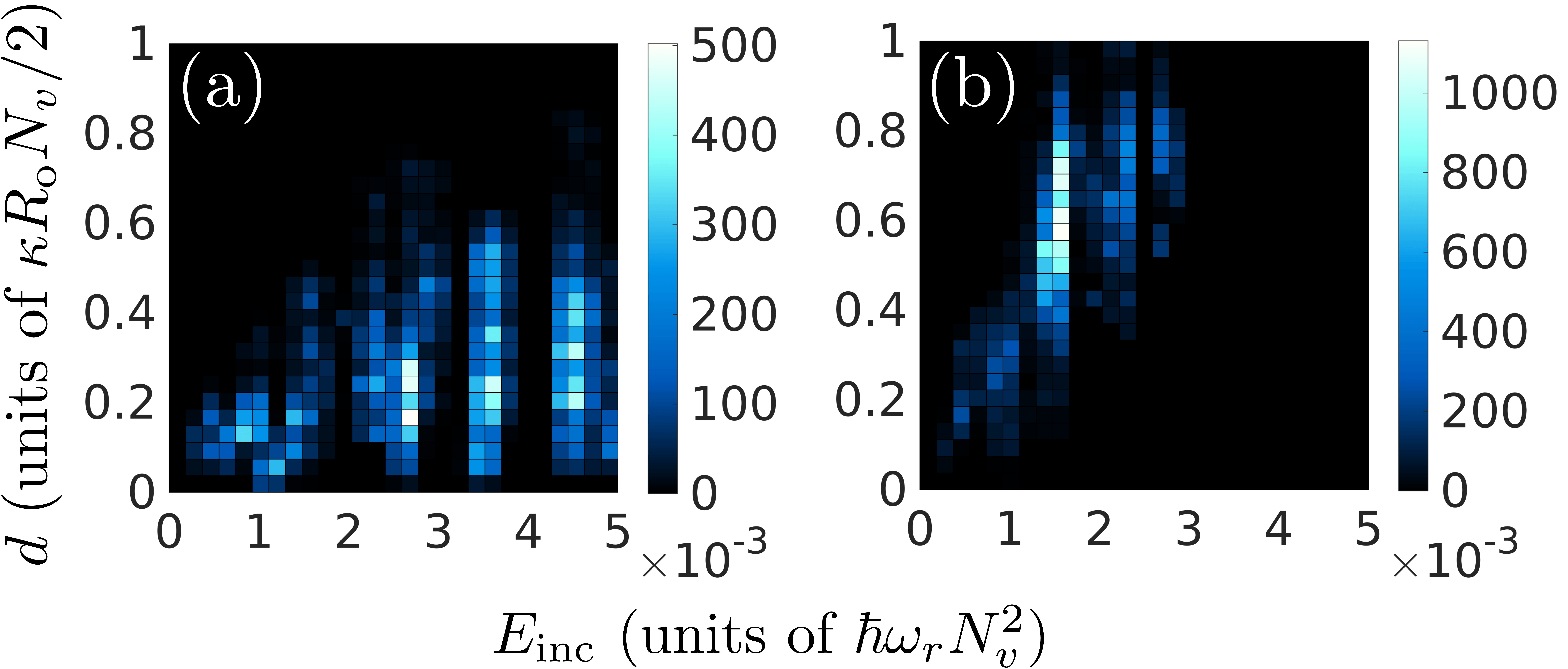}
\caption{Comparison of statistical behaviour between (a) the harmonic trap and (b) the uniform trap. For each dynamical simulation, the dipole moment of the vortex configuration is shown as a function of the incompressible kinetic energy per vortex number squared. The initial state in each plot is the bottom-left corner, and the evaporative heating increases the energy per vortex number squared over time. The data appears as columns because each vortex annihilation increases the energy per vortex number squared by a discrete amount.}
\label{fig:Eincstats}
\end{figure}

\begin{figure*}[t!]
\centering
\includegraphics[width=1.7\columnwidth]{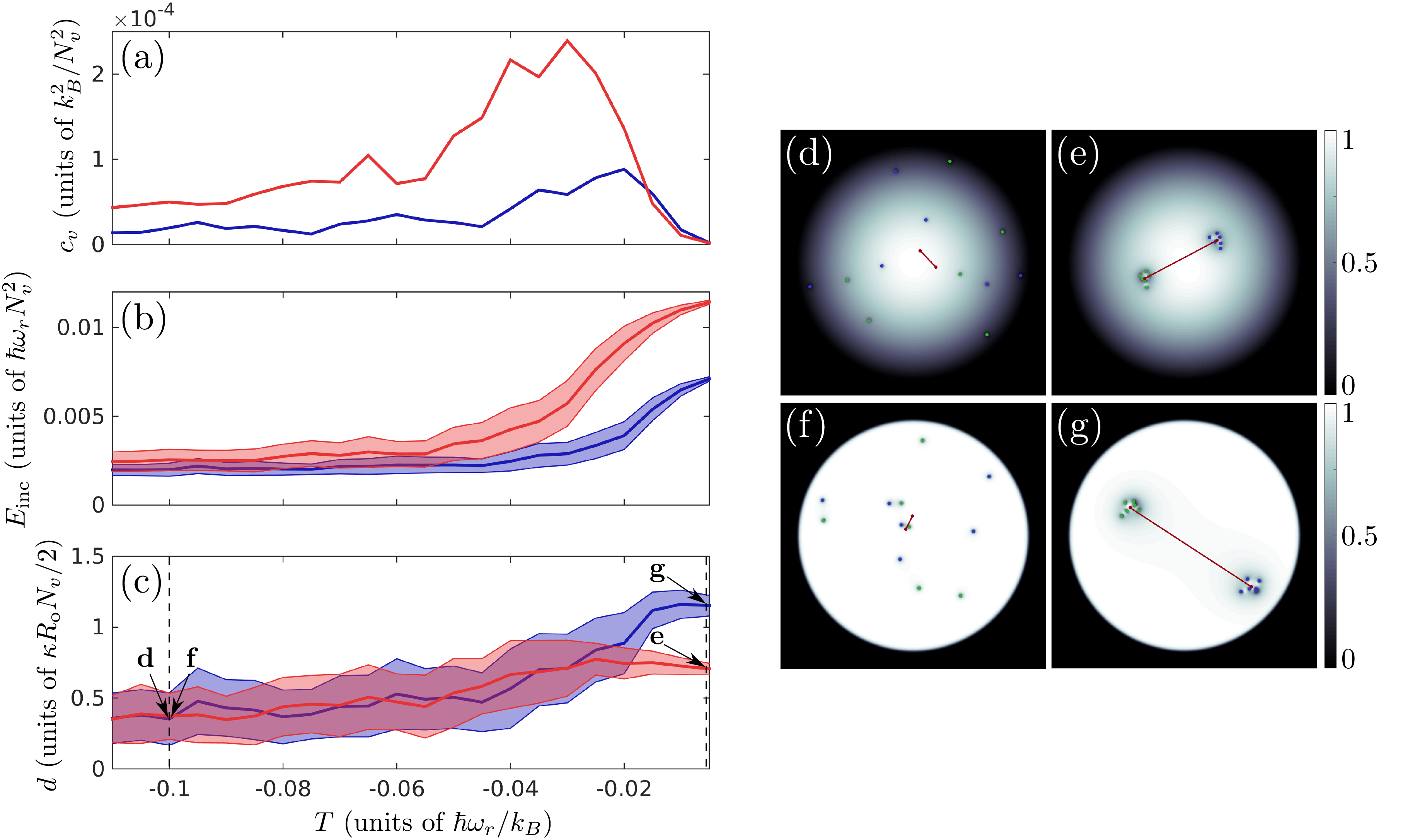}
\caption{Statistical data obtained from 100,000-step Markov Chain Monte Carlo simulations for a harmonic (red/light) and a uniform (blue/dark) trap for a total of 12 vortices with equal numbers of vortices and antivortices. The subfigures show (a) the specific heat, (b) the incompressible kinetic energy per vortex number squared, and (c) the dipole moment of the configuration, each plotted as a function of the statistical temperature. The shaded regions in (b) and (c) correspond to the standard deviation of each observable at a given temperature. The maximum in the specific heat indicates the transition to the Onsager vortex state in each trap, and is accompanied by an increase in both the energy per vortex number squared and the dipole moment. Frames (d) \& (e) and (f) \& (g) show typical vortex configurations at the temperature extremes in the harmonic and uniform traps, respectively, with labelling as in Fig.~\ref{fig:dynamics}. The temperatures shown in these frames are indicated in (c) with vertical dashed lines.}
\label{fig:MCstats}
\end{figure*}

Figure~\ref{fig:stats} shows little difference between the vortex number decay in the two traps. This suggests that the evaporation of vortices is only weakly affected by the details of the trapping potential. However, the dipole moment shows quantitatively different behaviour between the two traps, and indicates strongly enhanced clustering in the uniform trap. To better understand this difference, we construct a probability distribution of different vortex configurations generated by the dynamics in the space spanned by the dipole moment and energy per vortex number squared by taking the vortex configuration at each time-step to correspond to an independently sampled microstate. We choose to normalise the energy to the square of the vortex number to cancel out the $N_v^2$ scaling which occurs in the Onsager limit, as the system tends towards a multi-quantum vortex dipole configuration. Figure~\ref{fig:Eincstats} shows the resulting histograms for each trap. In the harmonic trap (a), the dipole moment shows no significant variation over the measured range of energy per vortex number squared, and hence there is no evidence that the system crosses the Onsager vortex transition. Conversely, the trend in the uniform trap (b) is a clear indication that the evaporative heating is on average increasing the dipole moment, causing the system to evolve towards the Onsager vortex state.

\subsubsection{Monte Carlo thermodynamics \label{ssub:stats2}}

In order to determine the statistical behaviour of the vortex gas beyond the range accessible via the dynamics, we implement a Markov Chain Monte Carlo (MCMC) algorithm for the two traps on a $256 \times 256$ grid. The algorithm is initialised by imprinting a random configuration of $N_v$ vortices into the condensate ground state using the imaginary time propagation method described in Sec.~\ref{sub:techniques}. We set $N_v=12$ (six vortices of each sign) to approximate the late time configurations of the dynamical simulations presented in Figs.~\ref{fig:dynamics} and \ref{fig:stats}. Keeping $N_v$ fixed, each step in the algorithm shifts a single randomly chosen vortex in the configuration and calculates the value of a predetermined weighting function $w$. This new configuration is then either accepted or rejected based on the change in the weighting function. Here, we use a Boltzmann factor $w = \exp(-E_{\textrm{inc}}/{k_{\rm B}}T)$ as our weighting function, defining $T$ as the statistical temperature of the vortex gas (which in this case is negative---see Refs.~\cite{kraichnan_two-dimensional_1980, simula_emergence_2014, viecelli_equilibrium_1995} for details). The choice of $T$ determines the most probable vortex configuration. Hence, we can vary $T$ to alter the statistical behaviour of the system---this is the basis of the evaporative heating interpretation of the dynamics. To characterise the temperature dependence, we measure three observables: the energy per vortex number squared, the dipole moment and the specific heat, defined as $c_v = \textrm{var}(E_{\textrm{inc}})/(N_v T)^2$. The system is evolved for 110,000 Monte Carlo steps, the first 10,000 of which are disregarded as the initial condition is, in general, unrepresentative of the chosen temperature. The results for both traps are shown in Fig.~\ref{fig:MCstats}. This MCMC data shows the transition from the disordered state to the Onsager vortex state in each trap, characterised most obviously by a maximum in the respective specific heat curves in Fig.~\ref{fig:MCstats}(a). In addition, both the energy per vortex number squared and dipole moment begin to rapidly climb around this critical temperature, signalling the formation of vortex clusters. For a uniform system with superfluid density $\rho_{\rm s}$, the critical temperature is predicted to be $T_c = -0.25 N_v \rho_{\rm s} \hbar^2 / m^2 k_B$ \cite{kraichnan_two-dimensional_1980, simula_emergence_2014}. For $N_v=12$ vortices, this yields a critical temperature of $T_c \approx -0.019\,\hbar \omega_r / k_B$, which agrees well with our data. In a harmonically trapped system, Fig.~\ref{fig:MCstats} shows that $T_c$ will be shifted towards lower temperatures compared to the uniform system.

The key differences between the two traps are evident in Fig.~\ref{fig:MCstats}. Figure~\ref{fig:MCstats}(c) shows that the dipole moment climbs to a significantly higher value at the highest temperatures in the uniform trap compared to the harmonic trap---the respective vortex configurations are displayed in frames (e) and (g). In fact, the dipole moment shows only a weak temperature dependence in the harmonic trap, the most marked effect being a decrease in its variance at high temperatures. This suggests that, even if the harmonically trapped system transition to the Onsager state, the resulting dipole moment would remain relatively small when compared to the steeper traps. Figure~\ref{fig:MCstats}(b) also shows that the energy per vortex number squared required to cross the transition is significantly higher in the harmonic trap. This provides further support for the absence of clustering in the GPE dynamics in the harmonic trap, as the evaporative heating does not supply enough energy to drive the system to these temperatures.

\begin{figure}[b]
\centering
\includegraphics[width=\columnwidth]{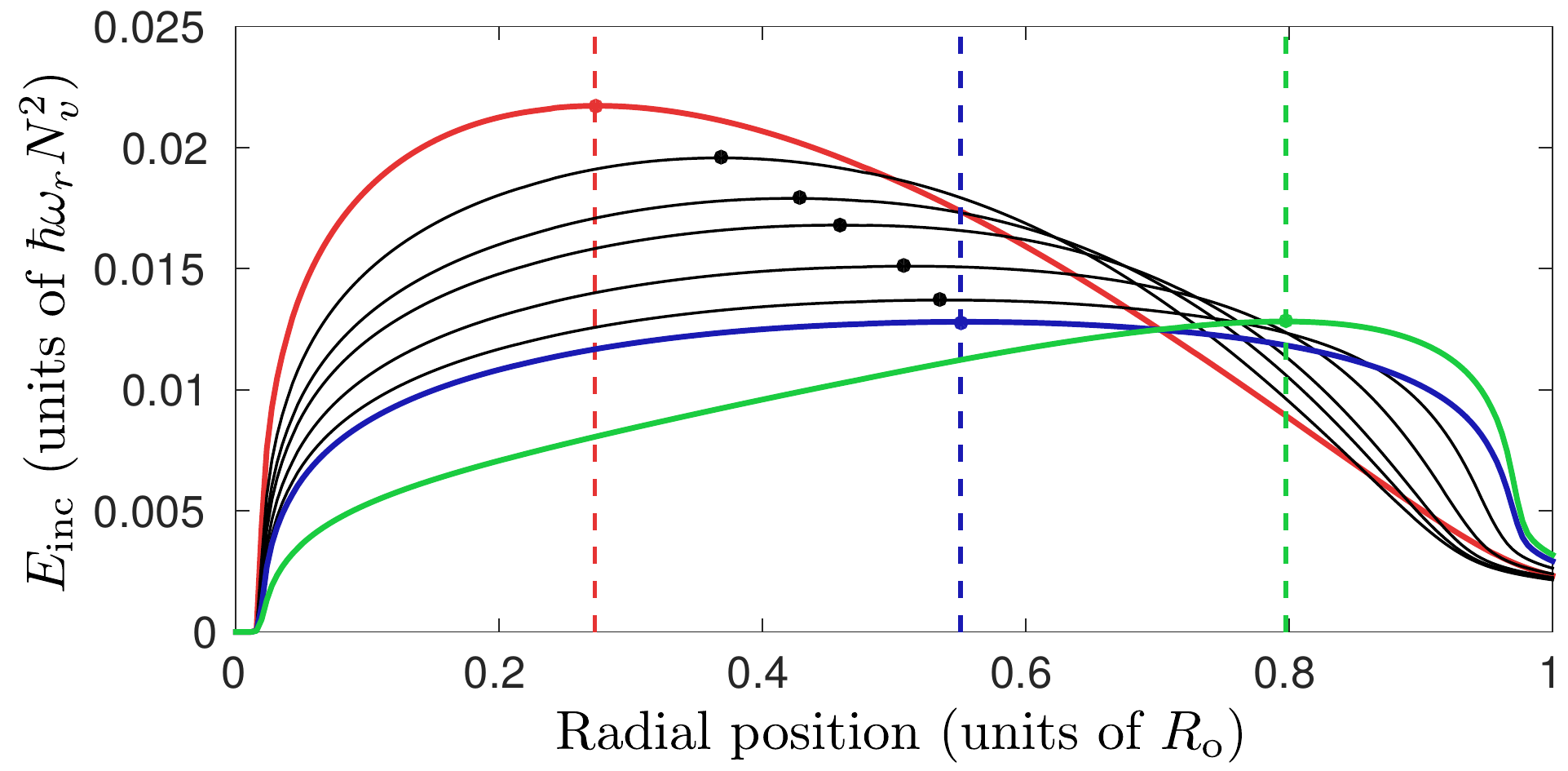}
\caption{Incompressible kinetic energies of a vortex--antivortex pair for a range of power-law traps computed using the GPE. The pair is placed symmetrically in the trap and both vortices are an equal radial distance from the center. In order of peak location from left to right, the power-law exponents are $\alpha=2$ (red), $\alpha=\lbrace 4,6,8,14,30 \rbrace$ (thin black lines) and $\alpha=100$ (blue). In addition, the dipole energy for an inverted trap (as described in the main text) is shown in light green (far right peak). The maximal separation is indicated on each curve with a circle, and is emphasised further on the two extreme power-law traps, as well as the inverted trap, with a vertical dashed line.}
\label{fig:Energyvsr}
\end{figure}

\subsubsection{Maximum achievable dipole moment \label{ssub:stats3}}

We can predict numerically the maximal separation of the two Onsager vortex clusters in a given system by calculating the energy of a vortex dipole as a function of the separation of the vortex and the antivortex. This yields further insight as to why the two traps show different clustering behaviour. In an infinite system, increasing the dipole separation will logarithmically increase the energy of the Onsager dipole without bound. However, for a bounded system, there exists a separation which maximises the energy. For a harmonic trap, this maximum energy configuration also corresponds to a stationary state \cite{freilich_real-time_2010, middelkamp_bifurcations_2010, mottonen_stationary_2005, kuopanportti_size_2011}. The dipole energy landscapes obtained for various trap steepnesses are presented in Fig.~\ref{fig:Energyvsr}, showing that the energy maximising separation increases as a function of the steepness. This result explains why the MCMC dipole moments in Fig.~\ref{fig:MCstats}(b) asymptote to different values in the high temperature limit, as the two systems reach their highest energy at differing cluster separations. In addition to various power law traps, Fig.~\ref{fig:Energyvsr} shows the energy in an `inverted' trap. This trap consists of a steep wall ($\alpha=100$) and an additional repulsive Gaussian potential with a width of $R_{\textrm{o}}/3$ which causes the condensate density to dip in the center. By pushing the fluid radially outwards, the energy maximising separation of a vortex dipole increases significantly, suggesting that an Onsager state in this trap should have an even greater dipole moment than that in the $\alpha=100$ trap. We have confirmed this prediction with a dynamical GPE simulation presented as Movie S3 in the Supplemental Materials \cite{supplement}.


\subsection{Vortex annihilation is a many-vortex process \label{sub:manyvortexann}}
The microscopic underpinning of the evaporative heating mechanism of vortices is vortex--antivortex annihilation \cite{simula_emergence_2014}. Scalar Bose--Einstein condensates with quantised vortices have two types of low-lying excitations---Bogoliubov phonons and vortex waves \cite{Simula2008b,Simula2010a,Simula2013a}. Such modes can resonate, mediating vortex--sound interactions \cite{leadbeater_sound_2001,zuccher_quantum_2012}. In principle such vortex--phonon interactions could cause vortex--antivortex pairs to annihilate via soundwave emission, which would account for the conservation of energy and momentum. However, for a single vortex--antivortex pair this does not occur as has been supported experimentally \cite{neely_observation_2010} and shown theoretically \cite{lucas_sound-induced_2014}. If such vortex--antivortex pair annhilations are forbidden, this raises the question of how the vortex number can decay over time as observed both in the simulations and experiments \cite{kwon_relaxation_2014}.

\begin{figure}[b]
\centering
\includegraphics[width=0.98\columnwidth]{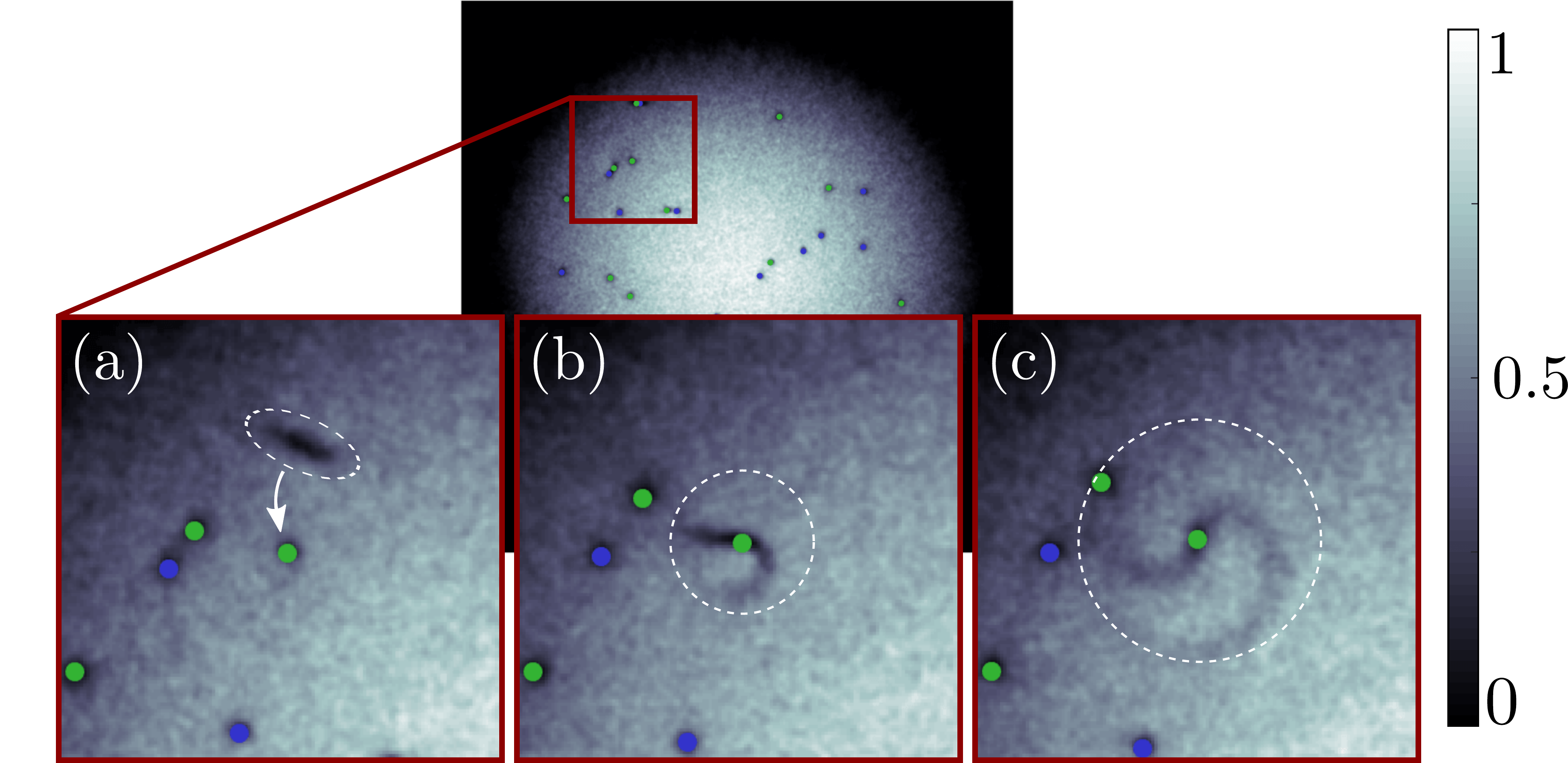}
\caption{A vortexonium state (formed from a vortex--antivortex pair) highlighted in (a) with a dashed oval colliding with an antivortex and dissipating into fluid soundwaves which disperse radially, indicated in (b) and (c) with dashed circles. The vortices and colorbar are labelled as per Fig.~\ref{fig:dynamics}. Supplemental Movie S4 shows the dynamics of this event \cite{supplement}, including the wavefunction phase.}
\label{fig:micro1}
\end{figure}

The answer must be that vortex--antivortex annihilation is a many-vortex process. Figure~\ref{fig:micro1} shows a three-vortex process whereby a vortex--antivortex pair has formed a neutral \textit{vortexonium} state (a rarefaction pulse), in which the individual vortex phase singularities are no longer discernible yet the excitation retains its identity as a spatially localised bound state. This excitation is reminiscent of positronium---a neutral bound state of an electron and a positron. The vortexonium, which is identifiable by a phase step, travels close to the speed of sound until it eventually scatters off an additional vortex or antivortex, as shown in Fig.~\ref{fig:micro1}(b) and (c). This decay process irreversibly disperses the energy and momentum of the vortexonium into sound waves \cite{nazarenko_freely_2006, smirnov_scattering_2015}. Until this secondary process occurs, the vortexonium can also re-form as a vortex--antivortex pair, an event which frequently occurs when a vortexonium state travels into the low density region near the boundary of the trap. The formation of vortexonium as a precursor to the vortex--antivortex annihilation process in 2D BECs has been discussed previously \cite{prabhakar_annihilation_2013, kwon_relaxation_2014, stagg_generation_2015, du_holographic_2015}. Here, we identify the three-vortex collision to be an essential part of the annihilation process in 2D superfluid turbulence.

\begin{figure}[b]
\centering
\includegraphics[width=0.85\columnwidth]{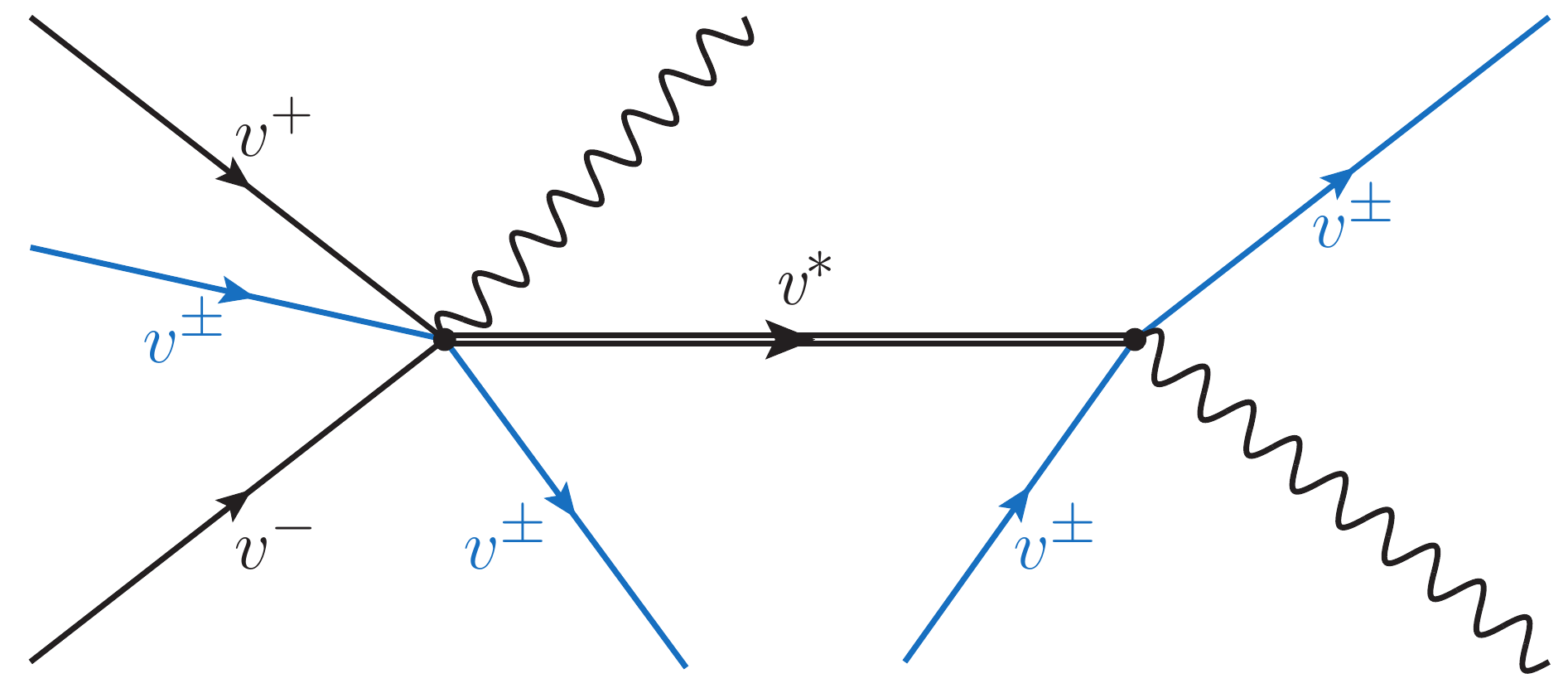}
\caption{Feynman diagram depicting the entire vortex--antivortex annihilation process observed, with time flowing from left to right. The straight lines represent vortices ($v^+$) and antivortices ($v^-$), the double line represents vortexonium ($v^*$), and the wavy line denotes the sound waves emitted at each vertex (the magnitude of the second burst of sound is far greater than the first). The light blue lines indicate participating catalyst vortices, which are not annihilated during the process.}
\label{fig:feynman}
\end{figure}

The question remains how these vortexonium states form to begin with. In a uniform system free from dissipation, an isolated vortex--antivortex pair will travel with constant velocity and inter-vortex separation. Therefore, some mechanism other than sound induced interaction must be responsible for reducing the pair's separation and forming vortexonium. In our simulations, we observe two ways this bound state can form. Firstly, a vortex--antivortex pair travelling towards a higher density region will reduce its separation in order to satisfy energy conservation, often forming a vortexonium state. However, this process only occurs in traps with shallow walls, where the density variation is significant. The second process we observe is the shrinking of a vortex--antivortex pair via a long-range interaction with a third vortex. By giving up some of its energy to this catalyst vortex, the pair can reduce their separation, ultimately resulting in a vortex--antivortex fusion event and the formation of a vortexonium state. We note that this latter process is ubiquitous in all traps studied. However, in the presence of dissipation, both the formation and annihilation of vortexonium would be possible without additional interactions, as the loss of energy would gradually drive vortex dipoles closer together regardless (see Sec.~\ref{sub:rateeqn}).

Combining these observations, we obtain a picture of the vortex--antivortex annihilation process, depicted as a Feynman diagram in Fig.~\ref{fig:feynman} which shows how four vortices are involved in the annihilation. Movie S4 in the Supplemental Materials \cite{supplement} shows one such four-vortex process. In the first stage a vortex--antivortex pair interacts with a catalyst vortex to produce a vortexonium state and in the second stage the vortexonium scatters off a catalyst vortex leading to the ultimate destruction of the vortex--antivortex pair and the emission of sound. The catalysts can be any vortex or antivortex in the system.


\subsection{Rate equation for evaporative heating of vortices \label{sub:rateeqn}}


Attempts have previously been made to fit a universal law to the vortex number decay \cite{kwon_relaxation_2014, stagg_generation_2015, schole_critical_2012, cidrim_controlled_2016}. Kwon \textit{et al.}~\cite{kwon_relaxation_2014} proposed a phenomenological model of the form $dN_v/dt = -\Gamma_1N_v - \Gamma_2N_v^2$, comprised of a linear term to model vortex drift out of the condensate and a nonlinear term to account for vortex--antivortex annihilation, where the $\Gamma_1$ and $\Gamma_2$ are the one-body and two-body decay constants, respectively.

We find that, due to the zero temperature of the GPE simulations, this equation does not provide an adequate fit to our vortex number decay curves. Instead, for $t \gtrsim 30 \, \omega_r^{-1}$, the vortex number decay is well described by a power law of the form $N_v(t) \propto (\omega_r t)^{-1/3}$ in all traps. This is evident in Fig.~\ref{fig:fitting}, which shows the number decay in a harmonic trap averaged over five simulations at $512 \times 512$ resolution. This power law was also observed by Schole \textit{et al.}~\cite{schole_critical_2012}, who further suggested that the vortex number rate equation should have the form $dN_v/dt \sim -N_v^4$. This would reflect the importance of a four-body loss process at zero temperature, in contrast to the one- and two-body loss observed in Kwon \textit{et al.}'s experiments \cite{kwon_relaxation_2014}. The four-vortex annihilation events discussed in Sec.~\ref{sub:manyvortexann} are consistent with this four-body loss mechanism.

To study the effect of the thermal cloud, we model non-zero temperatures using a damped Gross--Pitaevskii equation \cite{gardiner_stochastic_2001}:
\begin{align}
(i - \gamma) \hbar \frac{\partial}{\partial t} \psi(\textbf{r},t) = \bigg [ & \frac{-\hbar^2}{2m}\nabla^2 + V_{\textrm{trap}}(\textbf{r}) \nonumber \\
& + g_{\rm 2D}|\psi(\textbf{r},t)|^2 + i \gamma \mu \bigg ] \psi(\textbf{r},t),
\end{align}
where $\gamma$ is the temperature dependent dimensionless damping parameter, and $\mu$ is the chemical potential. We propose a general rate equation for vortex loss at all temperatures:
\begin{equation} \label{eq:Ndecay}
\frac{dN_v}{dt} = -\Gamma_1 N_v - \Gamma_2 N_v^2 - \Gamma_3 N_v^3 - \Gamma_4 N_v^4 - \ldots,
\end{equation}
where $\Gamma_n$ is the decay constant corresponding to a particular $n$-body loss mechanism. This model combines the one- and two-body loss processes observed in experiments \cite{kwon_relaxation_2014} with the higher order three- and four-vortex loss processes observed in our zero temperature simulations. Strictly, a three-vortex decay process is not possible since it would violate the vortex charge conservation law. We instead interpret the three-body term as the loss of two vortices arising from the collision of three (\textit{i.e.}~a vortexonium colliding with a catalyst vortex, as discussed in Sec.~\ref{sub:manyvortexann}).


We have chosen the damping parameter $\gamma=10^{-3}$ to study the vortex number decay behaviour at non-zero temperature. Figure~\ref{fig:fitting} shows the decay curves for zero temperature ($\gamma=0$) and non-zero temperature ($\gamma=10^{-3}$), each averaged over five simulations in a harmonic trap using a $512 \times 512$ numerical grid. We model both cases using Eq.~(\ref{eq:Ndecay}). For the $\gamma = 10^{-3}$ case, we find that the decay is best described by a one- and two-body model, with $\Gamma_1 = 0.14\,s^{-1}$, $\Gamma_2 = 0.044\,s^{-1}$ and $\Gamma_3 = \Gamma_4 = 0$. These values are in good agreement with those found by Kwon \textit{et al.}~\cite{kwon_relaxation_2014}. By contrast, the $\gamma=0$ case is best described by a three- and four-body decay model with decay constants $\Gamma_1 = \Gamma_2 = 0$, $\Gamma_3 = 1.2\times 10^{-4}\,s^{-1}$ and $\Gamma_4 = 8.1\times 10^{-7}\,s^{-1}$. We conclude that the three- and four-body vortex loss processes are characteristic of zero temperature systems, and that one- and two-body events become dominant at sufficiently high temperature. Quantifying the transition between these two behaviours at intermediate temperatures is left for future study.

\begin{figure}[h!]
\centering
\includegraphics[width=\columnwidth]{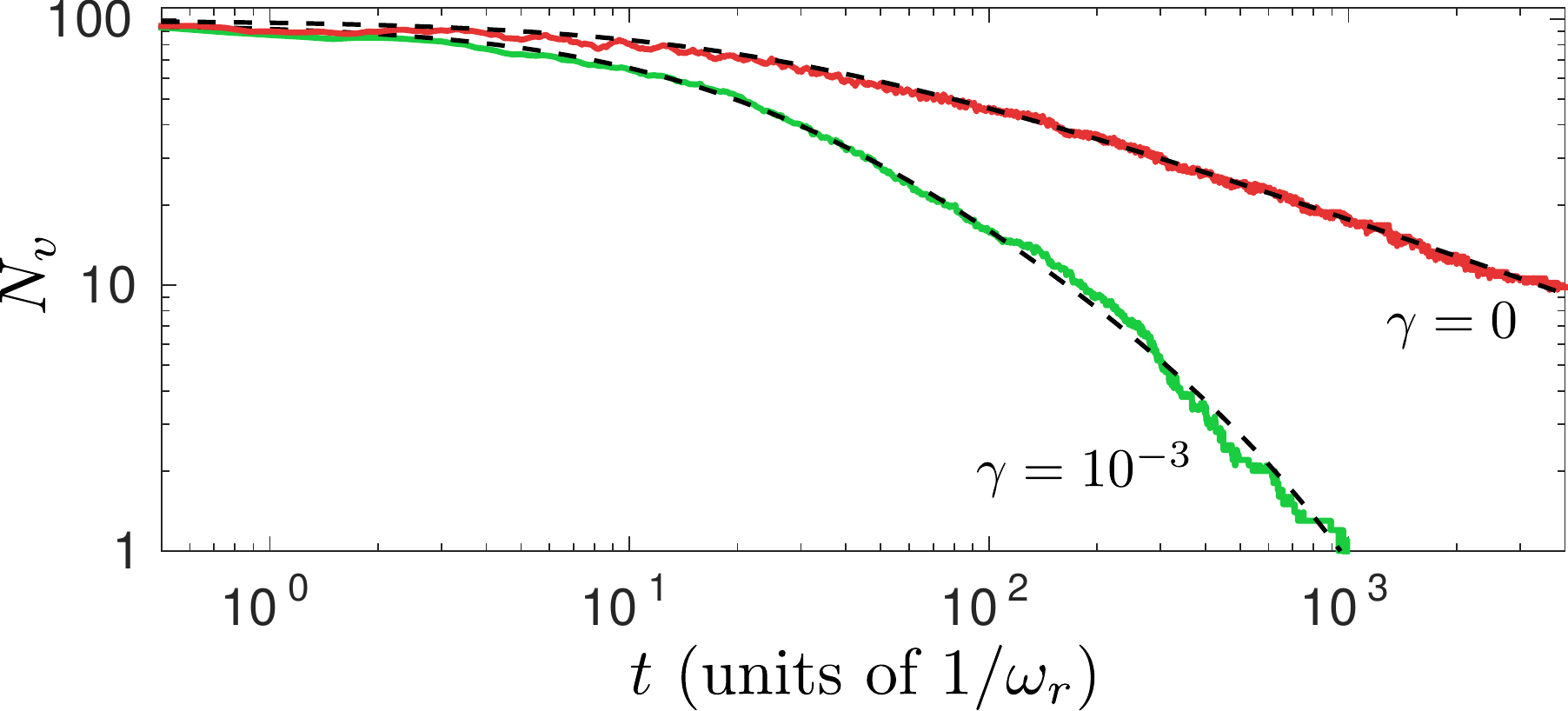}
\caption{Ensemble averaged vortex number decay curves for harmonically trapped systems at zero temperature ($\gamma=0$, solid red/dark line) and non-zero temperature ($\gamma=10^{-3}$, solid green/light line). The fits for each curve to Eq.~(\ref{eq:Ndecay}) are shown as black dashed lines, with $\Gamma_1 = 0.14\,s^{-1}$, $\Gamma_2 = 0.044\,s^{-1}$, $\Gamma_3 = \Gamma_4 = 0$ for the non-zero temperature case, and $\Gamma_1 = \Gamma_2 = 0$, $\Gamma_3 = 1.2\times 10^{-4}\,s^{-1}$, $\Gamma_4 = 8.1\times 10^{-7}\,s^{-1}$ for the zero temperature case.}
\label{fig:fitting}
\end{figure}


\subsection{Interaction between vortices and boundaries \label{sub:boundaryint}}

In our harmonic trap simulations, the multi-vortex collision process described in Sec.~\ref{sub:manyvortexann} is the only mechanism of vortex annihilation, excluding a small proportion of vortices which drift out of the condensate. By contrast, the presence of a hard boundary in the steeper traps allows for a number of additional phenomena relating to the dynamics and decay of vortex--antivortex pairs. In particular, we observe three distinct vortex--boundary collision processes, two of which give rise to additional vortex decay branches.

When a single vortex is near the boundary, it will pair up with its image vortex of opposite sign beyond the wall and travel around the circumference of the trap at high velocity. If the separation reduces sufficiently, this vortex--image pair can form a vortexonium with a phase step along the tangent of the wall. As this bound state travels around the boundary, it can either unbind and reform the vortex--image pair, or it can annihilate in much the same way as a vortexonium in the fluid bulk---by colliding with another vortex.

We observe a similar process involving the collision of a vortex--antivortex pair with the boundary. When the pair collides with the wall, it unbinds into two separate vortex--image pairs, which then travel around the boundary in opposite directions, as shown in Fig.~\ref{fig:micro2}(a)--(c). If travelling at high enough velocity, one or both of these new vortex--image pairs can form vortexonium excitations, which can then decay as described above. Often, the collision will be violent enough to cause one of the vortices in the initial pair to annihilate immediately, while the other one is left to travel around the boundary.

\begin{figure}[t]
\centering
\includegraphics[width=\columnwidth]{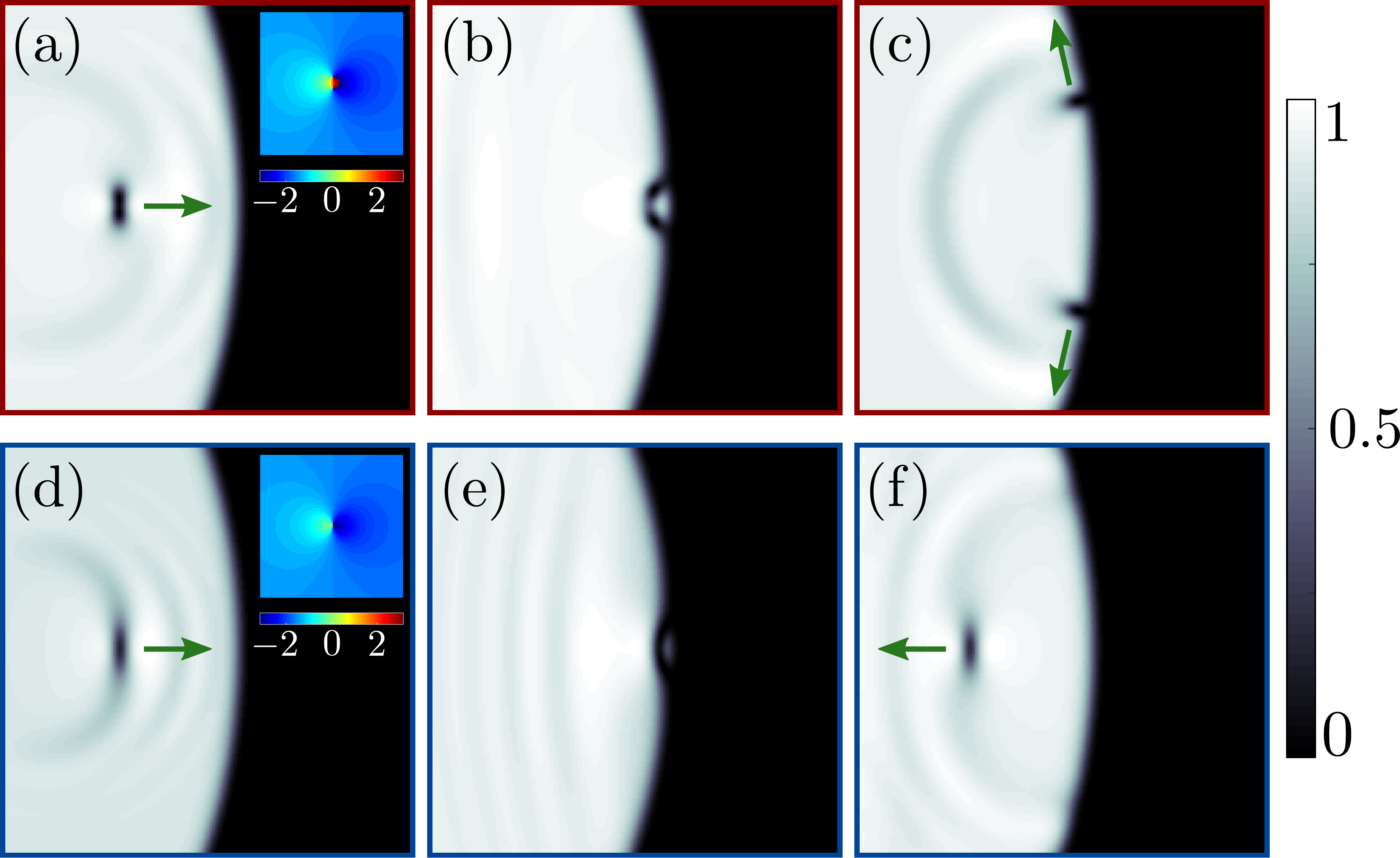}
\caption{(a)--(c) Unbinding of a vortex pair at the boundary in the uniform trap; (d)--(f) reflection of a vortexonium state at the boundary in the uniform trap. The green arrows show the direction each excitation is travelling. The insets in (a) and (d) show the phase of the wavefunction in the corresponding frame, showing the two singularities in (a) and the phase step in (d). The soundwave produced by each collision event propagates outwards in (c) and (f). The colorbar is normalised to the maximum condensate density, as in Fig.~\ref{fig:dynamics}. Movies S6 and S7 in the Supplemental Materials \cite{supplement} show each event in full.}
\label{fig:micro2}
\end{figure}

If the initial conditions are such that the vortex--antivortex pair which is incident on the boundary has already fused and formed a vortexonium excitation, the collision dynamics become markedly different. Figure~\ref{fig:micro2}(d)--(f) shows that the vortexonium will not separate at the boundary, but rather reflect from it, reversing its propagation direction. This effectively changes the sign of the vortices in the bound state, and can be understood as an exchange of locations with the image vortices beyond the boundary. Effectively, the image vortexonium travels into the condensate, while the real vortexonium leaves.

Remarkably, for the steepest potentials, the proportion of vortices annihilated at the boundary (\textit{i.e.}~via one of the first two processes described above) accounts for approximately half of the total vortex loss. Despite this clear spatial dependence of annihilation behaviour which is absent in the harmonic trap, the vortex number decays at the same rate (see Fig.~\ref{fig:stats}) and the evaporative heating does not appear to be any more efficient. It seems plausible that boundary annihilations would increase evaporative heating efficiency, as less energy should be lost per annihilation (as the energy of a vortex in the low density close to the system's boundary is less than in the fluid bulk), leaving more for the remaining vortices. However, we have found no strong evidence of this effect.


\subsection{Onsager vortex formation as a function of trap steepness \label{sub:varyingalpha}}

We repeated our Gross--Pitaevskii simulations of decaying turbulence for a number of trap steepnesses ranging between the two extremes examined in Sections~\ref{sub:macro} and \ref{sub:statinterp} by varying the value of $\alpha$ in Eq.~\ref{eq:Vtrap}. Five GPE simulations were performed in each of the chosen trap geometries using a $512 \times 512$ grid, and the dipole moment curves obtained for each steepness were combined by taking averages at each point in time. These averaged dipole moment curves are shown in Fig.~\ref{fig:alphadipole}. On average, a steeper trap produces a larger dipole moment and thus a greater separation of vortex charge. As predicted from energy considerations in Sec.~\ref{sub:statinterp}, an inverted trap produces even stronger clustering than any of the power-law traps. For the power-law traps, it appears that the clustering behaviour saturates beyond a steepness of $\alpha \approx 30$. The dipole moments in Fig.~\ref{fig:alphadipole} should be compared with their predicted maximum values shown in Fig.~\ref{fig:Energyvsr}.

\begin{figure}[h!]
\centering
\includegraphics[width=\columnwidth]{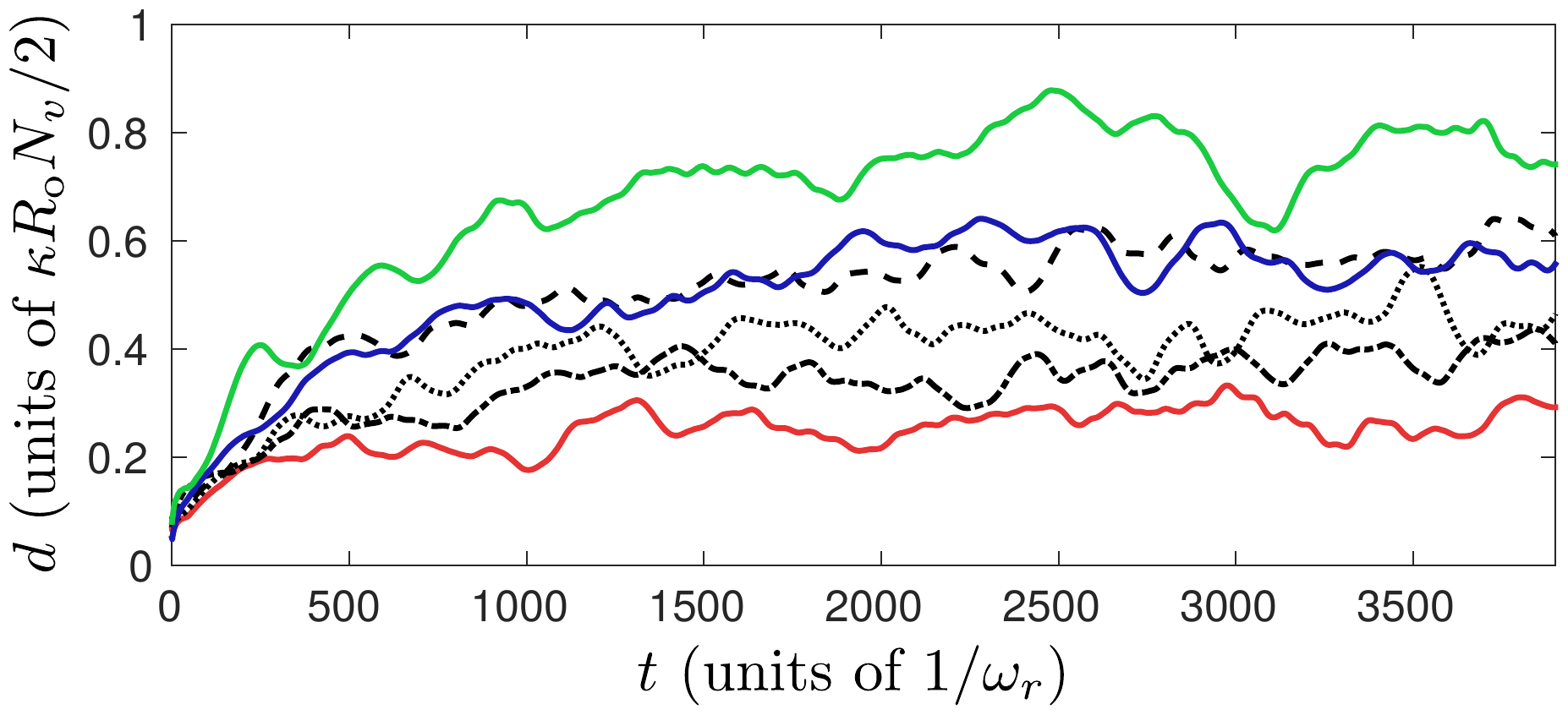}
\caption{Comparison of dipole moment evolution in traps of varying steepness. Each curve is averaged over five simulations. The trap steepnesses are (from bottom to top) $\alpha=2$ (solid red line), $\alpha=8$ (dot-dashed black line), $\alpha=14$ (dotted black line), $\alpha=30$ (dashed black line), $\alpha=100$ (solid blue line) and an $\alpha=100$ inverted trap (solid green line).}
\label{fig:alphadipole}
\end{figure}


\section{Discussion \label{sec:discussion}}

We have studied decaying two-dimensional quantum turbulence using the Gross--Pitaevskii model. We have considered Bose--Einstein condensates confined in generic power-law traps which, in particular, enables a comparison to be made between vortex dynamics in harmonically trapped condensates and in condensates confined in (nearly) uniform density disk traps. When an initial random vortex configuration is left to decay, we find that in uniform traps the vortices and antivortices arrange into Onsager vortex clusters due to the evaporative heating mechanism posited in Ref.~\cite{simula_emergence_2014}. However, when a harmonic trapping potential is used, the emergence of Onsager vortices is not obvious---a finding which agrees with experimental observations \cite{kwon_relaxation_2014}. To verify that these results are not specific to our randomly sampled initial vortex configurations, we repeated our simulations in both traps using a repulsive Gaussian laser potential to stir the fluid and produce the initial state vortex configuration, as in Ref. \cite{stagg_generation_2015}. Considering both lateral and circular stirring motions, the qualitative vortex evaporative heating behaviour in the harmonic and uniform traps was unaffected. This result was expected since a turbulent system should rapidly forget its history, washing out any initial state dependence.

We also performed Monte Carlo calculations to study equilibrium vortex configurations in harmonic and uniform traps. These calculations showed that the transition from disordered vortex configurations to the clustered Onsager vortex states exist also in harmonic traps but the resulting vortex dipole moment is significantly smaller than for uniform traps, which partly explains why the Onsager vortex clusters have not been observed to emerge in harmonically trapped Bose--Einstein condensates.

To obtain an improved understanding of the vortex evaporative heating mechanism \cite{simula_emergence_2014}, we carefully tracked the vortex--antivortex annihilation events in the simulations. At zero temperature, we found that vortex--antivortex pair annihilation in these quantum turbulent systems occurs via a combination of three- and four-body processes, involving one or two catalyst vortices, respectively, in addition to the annihilating pair. Firstly, a vortex-antivortex pair forms a \textit{vortexonium} bound state by either travelling through varying density (three-body) or interacting with a catalyst vortex (four-body). In both cases, this bound state then has to interact with a catalyst vortex for it to irreversibly decay into phonons. Indeed, it has been shown both experimentally \cite{neely_observation_2010} and theoretically \cite{lucas_sound-induced_2014} that an isolated vortex--antivortex pair is resistant to sound induced decay. By adding dissipation to the Gross-Pitaevskii model, we simulated a non-zero temperature system and found that the three- and four-body annihilation mechanisms become less important, and instead one- and two-body annihilation events begin to dominate, in agreement with experimental observations \cite{kwon_relaxation_2014}.

By considering power-law traps of varying steepnesses, we found that the vortex clustering tendency becomes stronger as the trap steepness is increased. Finally, we found that a locally and weakly anti-trapping potential \cite{Lewandowski2003a,Coddington2004a,Simula2005a} should provide the most promising route to experimental observation of the emergence of the Onsager vortices, which could possibly be detected using the vortex gyroscope imaging technique proposed in Ref.~\cite{powis2014a}.


\begin{acknowledgments}
We are grateful for useful discussions with Matthew Davis, Thomas Gasenzer and Andy Martin. We acknowledge support from an Australian Postgraduate Award (A. G.), the Australian Research Council via Discovery Project No. DP130102321 (T. S., K. H.) and the nVidia Hardware Grant Program (T. S.).
\end{acknowledgments}


\bibliography{groszek_onsager.bbl}
	
\end{document}